\documentclass[conference,compsoc]{IEEEtran}

\ifCLASSOPTIONcompsoc
  \usepackage[nocompress]{cite}
\else
  \usepackage{cite}
\fi

\ifCLASSINFOpdf
\else
\fi

\usepackage{array}

\usepackage{amssymb}%
\usepackage{pifont}%

\newcommand{\cmark}{\ding{51}}%
\newcommand{\xmark}{\ding{55}}%

\usepackage{graphicx}
\usepackage[dvipsnames]{xcolor}
\usepackage{xspace}
\usepackage{listings}
\PassOptionsToPackage{hyphens}{url}\usepackage[colorlinks = true,
              linkcolor={green!80!black},
  citecolor={red!70!black},
  urlcolor={blue!70!black}]{hyperref} %

\usepackage{cleveref}
\usepackage{booktabs}
\usepackage{graphicx}
\usepackage{tabularx}
\usepackage{multirow}

\definecolor{c1}{RGB}{207, 34, 46}
\definecolor{c2}{RGB}{5, 80 , 174}
\definecolor{c3}{RGB}{36, 41, 47}

\definecolor{codegreen}{rgb}{0,0.6,0}
\definecolor{codegray}{rgb}{0.5,0.5,0.5}
\definecolor{codepurple}{rgb}{0.58,0,0.82}
\definecolor{textcolor}{rgb}{0.3,0.3,0.3}
\usepackage{bold-extra}

\definecolor{backcolour}{rgb}{0.95,0.95,0.92}

\definecolor{rulecolor}{rgb}{0.87,0.87,0.87}

\lstdefinestyle{mystyle}{
    keywordstyle=\color{c2}\bfseries,
    numberstyle=\tiny\color{codegray},
    commentstyle=\color{gray},
    basicstyle=\linespread{1.3}\color{c3}\ttfamily\scriptsize\bfseries,
    breakatwhitespace=false,         
    breaklines=true,                 
    captionpos=b,                    
    keepspaces=true,                 
    numbers=left,                    
    numbersep=5pt,                  
    showspaces=false,                
    showstringspaces=false,
    showtabs=false,                  
    tabsize=2,
    frame=single,   
    rulecolor=\color{rulecolor},
    framexrightmargin=-2pt,
    framexbottommargin=0pt,
    framextopmargin=0pt,
}

\lstdefinelanguage
   [x64]{Assembler}     %
   [x86masm]{Assembler} %
   {morekeywords={CDQE,CQO,CMPSQ,CMPXCHG16B,JRCXZ,LODSQ,MOVSXD, %
                  POPFQ,PUSHFQ,SCASQ,STOSQ,IRETQ,RDTSCP,SWAPGS, %
                  rax,rdx,rcx,rbx,rsi,rdi,rsp,rbp, %
                  r8,r8d,r8w,r8b,r9,r9d,r9w,r9b, %
                  r10,r10d,r10w,r10b,r11,r11d,r11w,r11b, %
                  r12,r12d,r12w,r12b,r13,r13d,r13w,r13b, %
                  r14,r14d,r14w,r14b,r15,r15d,r15w,r15b,endbr64, movzwl,pushq, xorl, lretq, movq, rip},
                    sensitive=true,%
   alsoother={\$},%
   morecomment=[l]\;,%
   morecomment=[n]{\#=}{=\#},%
   morestring=[s]{"}{"},%
   morestring=[m]{'}{'},
                 }[keywords,comments,strings]%

\newcommand{\new}[1]{{#1}}

\newcommand{\vc}{\texttt{\#VC}\xspace}
\newcommand{\ve}{\texttt{\#VE}\xspace}
\newcommand{\vmmcall}{\texttt{vmmcall}\xspace}
\newcommand{\mov}{\texttt{mov}\xspace}
\newcommand{\rdtsc}{\texttt{rdtsc}\xspace}

\newcommand{\mmiowrite}{\texttt{mmio}\xspace}

\newcommand{\vmexit}{\texttt{vmexit}\xspace}

\newcommand{\rdx}{\texttt{rdx}\xspace}
\newcommand{\rax}{\texttt{rax}\xspace}
\newcommand{\rcx}{\texttt{rcx}\xspace}

\newcommand{\rip}{\texttt{rip}\xspace}

\newcommand{\deadbeef}{\texttt{0xdeadbeef}\xspace}

\newcommand{\exitreason}{\texttt{exit\_reason}\xspace}

\newcommand{\kin}{\texttt{k$_{in}$}\xspace}
\newcommand{\kapp}{\texttt{k$_{app}$}\xspace}
\newcommand{\khyp}{\texttt{k$_{hyp}$}\xspace}

\newcommand{\skipg}{$S$\xspace}
\newcommand{\raxr}{$R_{rax}$\xspace}
\newcommand{\raxrskip}{$R_{rax}S$\xspace}

\newcommand{\raxw}{$W_{rax}$\xspace}
\newcommand{\raxwskip}{$W_{rax}S$\xspace}

\newcommand{\readg}{$R_{mem}$\xspace}
\newcommand{\writeg}{$W_{mem}$\xspace}

\newcommand{\pkbase}{PK$_{base}$\xspace}
\newcommand{\pelast}{PE$_{last}$\xspace}
\newcommand{\pvbase}{VK$_{base}$\xspace}

\newcommand{\vaci}{VA$_{ci}$\xspace}

\newcommand{\codename}{{\sc WeSee}\xspace}

\newcommand{\pagec}{Page$_{c}$\xspace}
\newcommand{\pager}{Page$_{r}$\xspace}

\newcommand{\text}{\texttt{.text}\xspace}
\newcommand{\etal}{et al.\xspace}

\newcommand{\usermodehelper}{\texttt{call\_usermodehelper}\xspace}
\newcommand{\loc}{LoC\xspace}
\newcommand{\cvm}{VM\xspace}

\crefname{figure}{Fig.}{Fig.}
\crefname{section}{\S}{\S}
\crefname{appendix}{Appx.}{Appx.}
\crefname{table}{Tab.}{Tab.}
\crefname{listing}{Lst.}{Lst.}
\crefalias{section}{appendix}

\begin{document}

\title{\codename: Using Malicious \#VC Interrupts to Break AMD SEV-SNP\\
\thanks{Identify applicable funding agency here. If none, delete this.}}

 \author{
\IEEEauthorblockN{Benedict Schlüter \qquad Supraja Sridhara \qquad Andrin Bertschi \qquad Shweta Shinde}
\IEEEauthorblockA{ETH Zurich}
}

\maketitle

\begin{abstract}
AMD SEV-SNP offers VM-level trusted execution environments (TEEs) to protect the confidentiality and integrity for sensitive cloud workloads from untrusted  hypervisor controlled by the cloud provider.
AMD introduced a new exception, \vc, to facilitate the communication between the VM and the untrusted hypervisor. 
We present  \codename attack, where the hypervisor injects malicious \vc into a victim VM's CPU to compromise the security guarantees of AMD SEV-SNP. 
Specifically, \codename injects interrupt number 29, which delivers a \vc exception to the VM who then executes the corresponding handler that performs data and register copies between the VM and the hypervisor. 
\codename shows that using well-crafted \vc injections, the attacker can induce arbitrary behavior in the VM.
Our case-studies demonstrate that \codename can leak sensitive VM information (kTLS keys for NGINX), corrupt kernel data (firewall rules), and inject arbitrary code (launch a root shell from the kernel space).
\footnotetext[0\def\thefoornote{}]{This is the author's version of the IEEE S\&P 2024 paper.}

\end{abstract}

\IEEEpeerreviewmaketitle

\section{Introduction}

Hardware-based trusted execution environments (TEEs) make it possible to execute sensitive computation on an untrusted cloud while providing confidentiality and integrity guarantees. 
Hardware platforms and vendors, including Intel, AMD, Arm, and IBM have rolled out or announced support for VM-level TEEs~\cite{sev-snp,cca,ibm-cc,tdx}.
AMD Secure Nested Paging (SEV-SNP) provides both confidentiality and memory integrity of VM execution~\cite{sev-snp}. It is in production on major cloud service providers, including Azure, Google Cloud, and AWS~\cite{azure-sev,google-sev,aws-sev} and has been applied to 
security-sensitive workloads~\cite{ms-coco-ex, azure-container, amd-cvm-use-in-companies, oracle-cc-db, cc-summit}.

Since the cloud provider controls the hypervisor on a cloud platform, TEEs such as AMD SEV-SNP deem this privileged software to be untrusted. 
In the CVM setting, the hypervisor is still responsible for configuration and management of resources, including interrupts. 
This change in the trust model strengthens the security guarantees  
by enforcing that the hypervisor can no longer access the VM's memory or registers in plain text, thus protecting the VM. 
However, this breaks several traditional systems abstractions that are essential to execute VMs. 
For example, the VM needs services such as CPUID and hypercalls 
from the hypervisor, which is rendered impossible if the hypervisor cannot access the VM's state in unencrypted form. 

To address this challenge, AMD SEV-SNP introduces new interfaces between the untrusted hypervisor and the trusted VM.
This allows the VM to re-enable the essential functionality, while being able to control the use of the interface and perform sanitization and correctness checks. 
For example, the VM can selectively exchange data from its own memory to a shared memory region with the hypervisor. 
Further, to maintain performance and compatibility, AMD SEV introduces a new exception called {\em VMM Communication Exception
(\vc)}~\cite{vc-intro}. The CPU raises this exception when the VM needs to communicate with the hypervisor. This enables the VM's exception handler to perform the communication via the shared memory region transparently without changing the guest kernel or applications. 
Whenever the VM executes an instruction that require hypervisor intervention (e.g., \texttt{cpuid}, \texttt{rdtsc}, memory mapped I/O), the SEV-SNP hardware raises a \vc exception.

Our attack, called \codename, abuses the \vc exception to break the security guarantees of AMD SEV-SNP. 
Our first observation is that the hypervisor can inject a {\em malicious \vc} into a CPU that is executing a SEV-SNP VM at any time. 
Specifically, the hypervisor has the ability to inject external interrupts to the CPUs, including \vc which is yet another exception. 
Our second observation is that SEV-SNP invokes the \vc exception handler in the VM without checking the authenticity of the root cause. Specifically, the VC handler does not check if the VM indeed executed an instruction that would legitimately cause the CPU to generate a \vc exception. 
Our third observation is that the VC handler performs sensitive operations of copying data between the VM and the hypervisor to emulate the semantics of the instruction that generated the \vc. 
The handler is programmed to be bug-free and has checks to defend against Iago attacks, i.e., it  clears 
all registers and performs checks on the data values provided by the hypervisor before it uses them as per AMD specifications~\cite{amd-manual}. 
However, it is not programmed to defend against \vc that is maliciously injected by the hypervisor. 
Worse yet, each malicious \vc injection tricks the handler into emulating an instruction that either writes attacker-controlled data to the VM or leaks sensitive VM data to the hypervisor. 

\codename shows that with each malicious \vc injection from the hypervisor, the attacker can induce a basic primitive operation on the victim guest VM. For example, by faking a \vc for MMIO read, the attacker can achieve an arbitrary memory write---the hypervisor can write any value of its choice to any location in the victim VM. 
To achieve each basic primitive, we address several challenges such as ensuring that the victim VM does not crash (e.g., due to existing sanitization checks) and identifying particular execution points in the victim's execution to inject malicious \vc.
We demonstrate 4 main primitives namely: skipping instruction execution, leaking registers, corrupting registers, and arbitrary read/write to VM memory. 
 
Finally, \codename shows that the attacker can inject several malicious \vc to cascade the effect of the above basic primitives (e.g., memory write followed by memory read). We demonstrate several nuances that allow \codename to 
(a) inject consecutive \vc{s} before the VM resumes execution; (b) inject nested \vc{s} while the VM is executing the handler itself; and (c) combine consecutive and nested interrupts. 
Put together, this allows \codename to bring about highly expressive attacks such as arbitrary code injection and execution. 

We demonstrate the expressiveness of \codename with three end-to-end case studies. We leak kernel TLS session keys for NGINX with the arbitrary read. 
We use arbitrary write and code injection primitives to disable firewall rules and open a root shell. 
Orchestrating these case studies requires addressing challenges such as identifying suitable points of execution in the victim VM. Prior works have shown that AMD SEV-SNP is vulnerable to side channels that can be leveraged to achieve single-stepping primitives. However, \codename does not require such high-resolution information about the victim VM. Instead, we purely rely on the page fault sequences of the victim VM to perform our end-to-end attacks. 
We discuss potential software and hardware-based defenses to thwart \codename and argue for robust hardware mechanisms to limit the hypervisor's capabilities. 

Heckler~\cite{heckler-usenix} and \codename show that the hypervisor can abuse the notification mechanisms, existing and new respectively, to break CVM guarantees. These works points to a family of attacks called {
\em Ahoi attacks},\footnote{Ahoi is a signal word to call a ship or boat.
It is also an anagram of Iago~\cite{checkoway2013iago} with edit distance of one.}where the attacker sends malicious notifications, both in time and in value, to trick the victim.
Prior works that abuse timer interrupts and page faults can also be classified as Ahoi attacks, because they generate fake interrupts that allows the attacker to observe side-effects (e.g., cache and timing). However, \codename and Heckler generate interrupts that lead to explicit effect handler execution which directly update the global state (registers and memory) of the victim. 

In summary, we make the following novel contributions: 

\begin{itemize}
\item \codename abuses the \vc exceptions to break AMD SEV-SNP. 
\item \codename injects multiple well-crafted \vc exceptions into the victim VM to induce arbitrary reads, writes, and code injection. 
\item We demonstrate three case studies for \codename: leaking kTLS keys for NGINX, bypassing the firewall, and obtaining a root shell. 
\end{itemize}

We responsibly disclosed our findings to AMD on 26 October 2023 and the cloud providers on 5 February 2024. \codename was assigned CVE-2024-25742.

\noindent 
\codename tooling and PoC exploits are open-source at: \\ \url{https://ahoi-attacks.github.io/wesee}

\section{Overview}
\label{sec:overview}

AMD SEV-SNP disallows hypervisors from accessing guest VM registers and memory, thus necessitating \vc. 

\subsection{Background}
\label{ssec:background}

AMD virtualization extensions (AMD-V), introduced in 2006, provide hardware support for the hypervisor to create and launch guest VMs. 
Since the VMs cannot directly access certain system resources (e.g., \rdtsc), 
AMD-V allows the hypervisor to set up intercepts on particular instructions to manage and facilitate execution in the VMs.

\noindent{\bf Instruction interception for virtualization.}
Several operations in the guest VM---reads/writes to hypervisor-controlled Model Specific Registers (MSRs) and memory-mapped devices, accessing \texttt{rdtsc}, generic calls to the hypervisor---require co-operation with the hypervisor. 
When a VM executes such an instruction, the CPU intercepts it and automatically triggers a \vmexit that is handled by the hypervisor. 
Therefore, the CPU's instruction intercept mechanism facilitates calls to the hypervisor when these operations are performed in the guest VM. 
This mechanism is fast and transparent to the guest VM as it does not require the involvement of the guest kernel to support it. 
Further, the CPU maintains a fixed set of instructions for which it triggers a \vmexit, that the hypervisor can handle~\cite{amd-manual}.
For example, a guest VM uses the \vmmcall instruction to explicitly communicate with the hypervisor. 
First, the guest VM sets up registers and memory that contain values to indicate the reason to the hypervisor to process the VM's request.
Then, the guest VM executes the instruction (e.g., \vmmcall) that causes a \vmexit to the hypervisor. 
Due to the \vmexit, the hypervisor's handler is invoked where it can directly read the registers and memory from the guest VM and process the request. 
Since the hypervisor is trusted in AMD-V, this mechanism allows the VMs to execute unchanged while the hypervisor handles the special instructions.

\noindent{\bf AMD SEV-SNP.}
AMD SEV-ES and its successor SEV-SNP enable the creation and isolation of confidential VMs in various cloud deployments~\cite{amd-cvm-use-in-companies}. 
In particular, AMD SEV-SNP  provides a hardware-based trusted execution environment that protects the memory and registers of the VMs and renders them inaccessible to the hypervisor.  
This protection breaks some existing abstractions (e.g., instruction intercepts set by hypervisor). 
However, the isolated VM still needs to communicate with the hypervisor to perform different operations.
To address this gap, SEV defines a strict protocol to share data between the VM and the hypervisor using an unprotected shared memory region called the Guest Hypervisor Communication Block (GHCB)~\cite{ghcb-spec}. 

\subsection{Implications of Instruction Interception}
\label{ssec:vc-security}
\new{AMD adds support for secure instruction interception for \cvm{s}.\footnote{\cvm is a shorthand notation for SEV-SNP VM unless stated otherwise.} We analyze the need for such interception and the security implications of this support in AMD SEV-SNP.}

\begin{figure}
    \centering
    \includegraphics[scale=0.65]{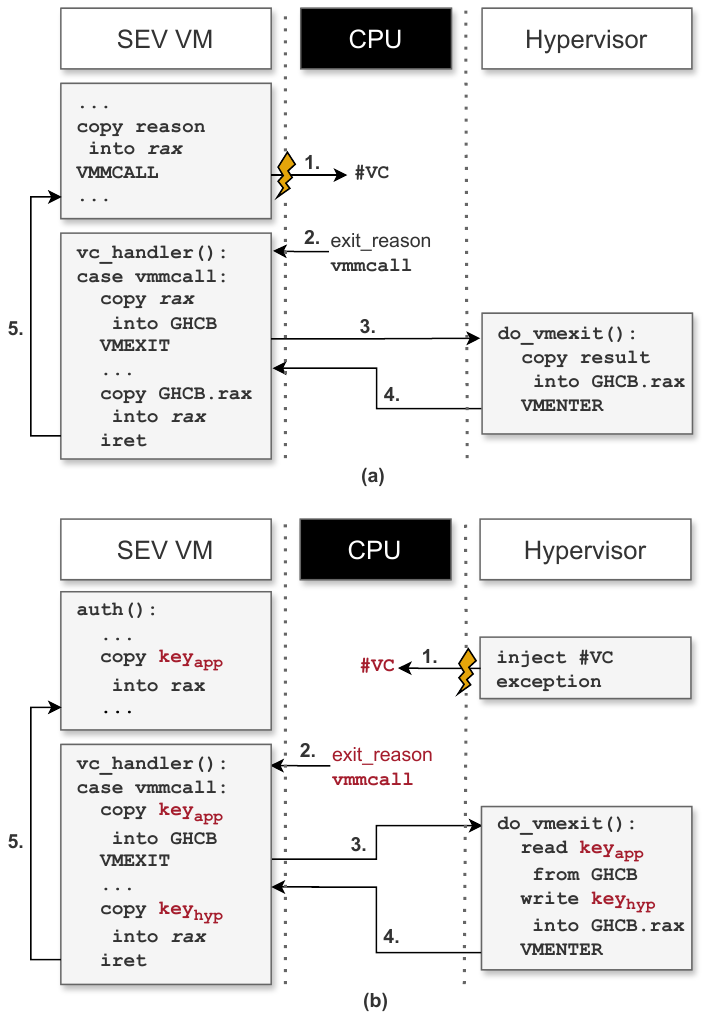}
    \caption{(a) Benign execution of \vc in SEV VM. 1. application sets up \rax and does \vmmcall 2. CPU intercepts the instruction and raises \vc with \exitreason set 3. VC handler copies values (\rax) into GHCB and exits to hypervisor 4. Hypervisor returns back to the VC handler which copies values (\rax) from GHCB 5. Return to application that caused \vc. (b) Attack execution with \codename.}
    \label{fig:vc-exception}
\end{figure}

\noindent{\bf Need for \vc exception.}
When AMD SEV-SNP is enabled, the hypervisor cannot directly access the VM registers and memory. 
Therefore, to enable instruction intercepts, VM's data needs to be copied into the GHCB. 
However, this copy to the GHCB is not performed automatically when the CPU intercepts an instruction.
There are several approaches to solving this issue: 
In the first approach, the user application in the VM is aware that it is executing in a \cvm and it explicitly uses the GHCB APIs by calling the guest kernel to copy data to the GHCB before exiting to the hypervisor. 
This approach requires invasive changes to the application code and breaks compatibility.
In the second approach, the guest kernel intercepts the instructions and performs the GHCB API calls. 
This solution is cumbersome as the guest kernel has no mechanism to determine which instructions in the applications to intercept (e.g., \texttt{mov} instructions executed for MMIO). 
Therefore, this approach either needs instrumentation of the application and the kernel or invasive changes to the application. 
There is a third approach: if the \cvm executes a specific set of instructions, the CPU raises a special exception that is delivered to the \cvm's kernel. 
The \cvm can register a handler for this exception where it can assess the reason for the exception and perform data copies to the GHCB, such that the VM can communicate with the hypervisor by performing a \vmexit in the exception handler. 
The hypervisor can then, in its own handler, access the GHCB, perform the relevant operations, and save the results in the GHCB. When the hypervisor returns control to the VM, the guest VM kernel handler uses the results from the GHCB and returns them to the user. 
If the hardware supports a new exception, this is the best approach.  
It is transparent to the user while the guest kernel can perform instruction-specific data copies as well as enforce interface sanitization. 
Therefore, SEV introduces a new exception called {\em VMM Communication Exception (VC)} with interrupt number 29, to facilitate \cvm and hypervisor communication.
The AMD SEV hardware has added support to intercept a certain set of instructions from a \cvm. When such an instruction is executed in the guest VM (user or kernel-space), the CPU generates a \vc exception and sets the \exitreason register to indicate the instruction that caused the exception. 
The rest of the execution flow is software-based via the VC handler in the guest and the \vmexit handler in the hypervisor where they use the GHCB as a shared memory region according to AMD's specification. \cref{fig:vc-exception}(a) depicts this and shows the pseudo-code and execution flow for \vc handling.

\noindent{\bf Example: Supporting \vmmcall with \vc.}
Consider an application in the \cvm that executes a \vmmcall to request data from the hypervisor, as shown in~\cref{fig:vc-exception}(a). 
When the \cvm executes a \vmmcall, i.e., instruction that should be intercepted by the hypervisor, the CPU triggers a \vc exception (Step 1). 
Then the CPU sets a hardware register \exitreason with the instruction (\vmmcall) and raises a \vc. 
As in the non-confidential case, the application in the \cvm still sets up the \rax register with the reason for the \vmmcall for the hypervisor to process the request. 
When the \vc occurs, the execution returns back to the VC handler in the \cvm (Step 2). 
 The VC handler is responsible for copying only the data required to process the \vmmcall into the GHCB based on the \exitreason register. 
For the \vmmcall, the handler only copies \rax into the GHCB. 
Therefore, it ensures that the \vc exception does not leak information to the hypervisor. 
Finally, the handler does a \vmexit to return control to the hypervisor to process the \cvm's request (Step 3). 
Before resuming the \cvm execution, the hypervisor writes the result of the \vmmcall in \rax of the GHCB and returns execution back to the VC handler (Step 4).
Then, the VC handler in the \cvm only copies the data requested based on the \exitreason register into the application context. 
For the \vmmcall, it only copies the \rax register back into the application. 
Finally, the application that caused the \vc can resume execution (Step 5). 
This process ensures that the hypervisor intercepts function correctly without compromising the \cvm's security.

\noindent{\bf Interrupt delivery to SEV-SNP VMs.}
The hypervisor manages the delivery of interrupts for the VMs. 
It can inject physical (e.g., timer interrupts) and virtual interrupts (e.g., virtio interrupts) to the VMs using the Interrupt Controller. 
x86-64 treats interrupt vector numbers 0-31 as exceptions~\cite{intel-sdm, amd-manual}.
Therefore, the hypervisor can use the interrupt controller to also inject the newly introduced \vc at any time to any cores that are executing the \cvm.

\subsection{\codename Attack}
\label{ssec:intfail-attack}
An attacker with the ability to inject interrupts into the \cvm (e.g., malicious hypervisor) can trigger a \vc at any time during the VM execution. 
Further, the hypervisor can write any value to the \exitreason register. 
More importantly, this interrupt causes the guest to execute the VC handler that examines the \exitreason, copies data from the VM to GHCB, and then does a \vmexit to the hypervisor who can look at the data in the GHCB. 
Furthermore, the VC handler also takes in the data provided by the hypervisor and copies it from the GHCB into VM memory. 
\new{In other words, the hypervisor can misuse \vc to execute the \cvm's handler at any point during the \cvm's execution and compromise it.} 
Using this insight, we present \codename attack that exploits the \vc to induce register and memory copy operations between the malicious hypervisor and the victim VM. 

\noindent{\bf Example.}
Consider a victim application executing in a \cvm. It authenticates a remote user by comparing an input key (\kin) with a secret key (\kapp). \cref{lst:example} shows this simple logic and~\cref{fig:vc-exception}(b) shows \codename attack on \cref{lst:example}.

\begin{lstlisting}[language={[x64]Assembler}, label={lst:example}, caption=Example application]
mov rbx, $kin
mov rax, $kapp
cmp rax, rbx
jne deny
auth: ...
jmp fin
deny: ...
\end{lstlisting}

The malicious hypervisor does not know the value of \kapp 
and therefore is not authenticated to use the application. 
With \codename, the hypervisor can successfully authenticate itself, thus breaking the SEV-SNP guarantees. 
The application copies the value of \kapp into \rax (Line $2$ in~\cref{lst:example}) before comparing it with \kin (Line $3$). 
The hypervisor maliciously injects the \vc exception after Line $2$.
\cref{fig:vc-exception}(b) Step 1 shows that the hypervisor can set the \exitreason as \vmmcall. 
This triggers the VC handler in the \cvm, which copies the \kapp from \rax into the GHCB and exits to the hypervisor (Step 2, 3). 
After this point \kapp is leaked to the hypervisor allowing it to authenticate successfully using this input. 
Alternatively, with this attack, the hypervisor can also write any value of its choosing (\khyp) into the GHCB's \rax to authenticate successfully. 
Specifically, it writes its own key which it sent to the application (\kin) to the GHCB before returning to the \cvm's VC handler (Step 4). 
The VC handler then copies the value of \rax to the application and returns to it (Step 5). 
This changes the value of \rax in the application leading to Line $3$ computing \kin==\xspace \khyp which will always be true as both the values are controlled by the hypervisor. 
Therefore, the hypervisor authenticates successfully and the  \texttt{auth} block is executed (Line $5$). 
Our example shows how a malicious hypervisor can use \codename to compromise the execution integrity as well as data confidentiality and integrity of the \cvm, thus breaking AMD SEV-SNP. 

\section{\codename Overview}

The VC handler executes in a VM that the hypervisor cannot modify or tamper with. 
To assess the potential of \codename, we manually analyze the Linux kernel v6.7-rc4 that implements the VC handler.
Our analysis shows that there are constraints on: 
(a) when the hypervisor can trigger certain \vc (e.g., MMIO can only be triggered on a \mov instruction); and (b) what operations the handler performs based on \exitreason (e.g., \vmmcall can only read and write \rax). 
Therefore, to mount \codename attacks, the hypervisor has to ensure that it achieves its desirable effects (e.g., change the value of \rax to \deadbeef) without resulting in a VM crash. 
More importantly, the \vc handler performs several other operations (e.g., masking registers before copying to GHCB) and checks (e.g., checking the operands of the instruction that caused the \vc) that either hinder the attacker from achieving its desired effect by causing check failures or result in excess execution  (i.e., undesirable effects).
Thus, we have to carefully craft the \exitreason at a well-chosen execution point in the VM such that we precisely induce the desirable effects of \vc handling while avoiding any checks and undesirable effects. 
\codename aims to leverage the hypervisors' ability to inject multiple \vc{s} to cascade the desired effects of the handler to bring about expressive attacks. For example, use one \vc to change the \rax value and then use another \vc to change \rcx value. 
While injecting multiple \vc{s} is straightforward, doing so requires 
knowing when to inject the next \vc while 
ensuring that the handler does not crash (e.g., due to nested exceptions beyond a certain depth).

\subsection{Analysis of \#VC Handler}

\begin{table*}
\caption{Intercept Events with \#VC handler implementation in SNP Linux guest. \\ \textup{*: Register is masked before/after exchanged with the hypervisor, \#reg: Register depends on the register used in the assembly instruction, GPA: Guest Physical Address, \texttt{vmgexit}: Guest exits to the hypervisor.} }
\label{tab:vc-handler}
\centering
\scriptsize

\begin{tabular}{lllllcc}
\toprule
Event         & Description              & \begin{tabular}[c]{@{}l@{}}Reg. copied\\to Hyp\end{tabular} & \begin{tabular}[c]{@{}l@{}}Reg. copied \\from Hyp\end{tabular}  & \begin{tabular}[c]{@{}l@{}}Sample Instr.\end{tabular}& \begin{tabular}[r]{@{}l@{}} \texttt{vmgexit} \end{tabular} & \begin{tabular}[c]{@{}l@{}}Used in\\ \codename \end{tabular}  \\
\midrule

NPF MMIO Read     & Memory mapped I/O read          & -                                                   & \#reg                                                      & mov rbx, [rax]   & \cmark & \cmark                                       \\

NPF MMIO Write    & Memory mapped I/O write         & \#reg                                                   & -                                                         & mov [rax], rbx    & \cmark & \cmark                                         \\

VMMCALL           & Call to VM Monitor              & rax, cs*                                            & rax                                                                                                     & vmmcall   & \cmark     & \cmark                                           \\

RDTSC/ RDTSCP     & Read Time Stamp Counter         & -                                                   & rax, rdx, rcx                                                                                             & rdtsc    & \cmark      & \xmark                                           \\

RDPMC             & Read Perf. Monitor Counter      & rcx                                                 & rax, rdx                                                                                              & rdpmc    & \cmark               & \xmark                                  \\

RDMSR             & Read from MSR                   & rcx                                                 & rax, rdx                                                                                             & rdmsr    & \cmark              & \xmark                                   \\

WRMSR             & Write to MSR                    & rcx, rax, rdx                                       & -                                                                                                      & wrmsr    & \cmark                         & \xmark                       \\

CPUID             & CPU Identification              & rax*, rcx*, xcr0*                                   & rax, rbx, rcx, rdx                                        & cpuid      & \cmark & \xmark    \\

IOIO\_PROT   &  IO Ports (IN, OUT, INS, OUTS)      & rax*    & rax*     &   in eax, 0   & \cmark & \xmark \\

DR7 write         & Debug Control Reg. write        & \#reg*                                                & -                                                                                                        & mov, dr7, rax   & \cmark        & \xmark                                  \\

RD7 read          & Debug Control Reg. read         & -                                                   & -                                                                                                      & mov rax, dr7  & \xmark    & \xmark                                       \\

INVD              & Invalidate Internal Caches      & -                                                   & -                                                                                                       & invd  & \xmark         & \xmark                                          \\

WBINVD            & Write Back and Invalidate Cache & -                                                   & -                                                                                                     & wbindv    &     \cmark          & \xmark                                 \\

MONITOR/ MONITORX & Set Up Monitor Address          & -                                                   & -                                                                                                      & monitor  & \xmark         & \xmark                                        \\

MWAIT/ MWAITX     & Monitor Wait                    & -                                                   & -                                                                                                         & mwait    & \xmark         & \xmark                                       \\

AC                & Alignment Check          & -                                                   & -                                                                                                       & mov [0x1001], rax  & \xmark & \xmark \\

\bottomrule
\end{tabular}

\end{table*}

The \cvm's VC handler copies different registers and memory to and from the GHCB, based on the intercepted instruction indicated by the \exitreason register. 
We analyzed the VC handler in the latest version of the Linux kernel (v6.7-rc4 as of this writing). It is implemented as per the GHCB protocol specified by AMD SEV-SNP developer manual \cite{amd-manual} and handles $19$ instruction intercept events. 
Out of these, we found that  $10$ events lead to register and memory copies (see~\cref{tab:vc-handler}). 
For example, in case of \texttt{rdpmc} the handler sends values via register \rcx to the hypervisor, and copies values from the hypervisor into registers \rax and \rdx.
For the remaining $9$ events, the VC handler does not send or receive any data from the hypervisor. 
Of the 10 events, in \codename, we only use the $3$ events shown in~\cref{tab:vc-handler} and show that they are sufficient to build strong attack primitives such as arbitrary register reads/writes, memory reads/writes, code injection.

\begin{figure}
    \centering
    \includegraphics[scale=0.8]{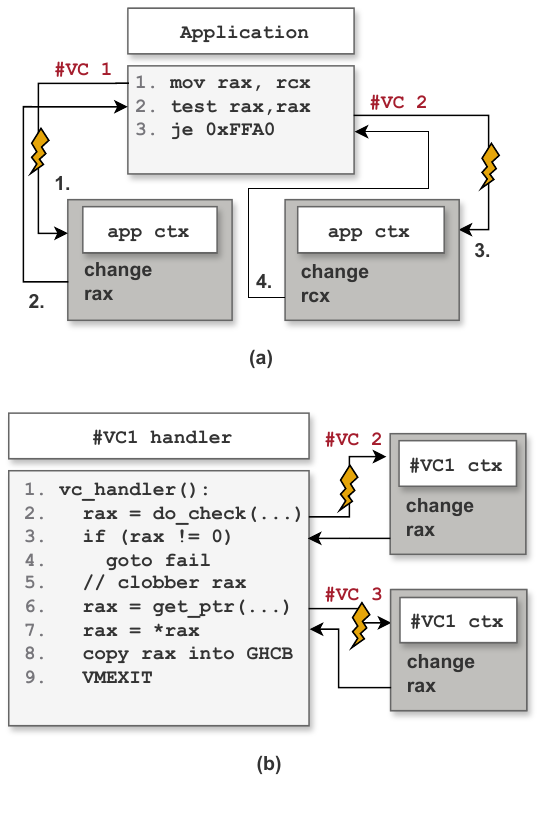}
    \caption{(a) VC chaining. VC1 changes \rax, VC2 changes \rcx. (b) 1-level VC nesting.}
    \label{fig:vc-nesting}
\end{figure}

\noindent{\bf Chaining multiple \vc.}
AMD SEV-SNP allows the hypervisor to inject consecutive \vc{s} to the same CPU. 
Consider a case where the \cvm is executing a user program. The hypervisor injects a first \vc and waits till the \cvm's kernel executes the corresponding handler. When the \cvm's kernel returns from the handler, the hypervisor can inject a second \vc right before the user program resumes on the CPU.
This execution flow is feasible because each \vc handler sets up its own stack and tears it down before returning. Thus, the hypervisor can inject two consecutive \vc{s} and use them to perform two different changes to the victim VM. 
For example, ~\cref{fig:vc-nesting} (a) shows how the hypervisor can change \rax and then \rcx by chaining two \vc{s}. 
The hypervisor can also inject \vc{s} in a staggered fashion, allowing the application to execute a few instructions between the first and the second \vc{s}.
In our analysis, we did not find hardware or software limitations on the number of \vc{s} that a hypervisor can inject consecutively or with staggering.

\noindent{\bf Nesting \vc in non-critical section.}
The hypervisor can inject a \vc while the guest kernel is executing a \vc handler, thus causing nested interrupts. This is functionally safe because each \vc sets up its own stack. More importantly, from an attack perspective recall that a \vc handler changes the state (register or memory) of the code that was executing when the \vc was injected. 
In case of a nested interrupt, the effects of the second \vc handler change the state of the first \vc handler. As shown in ~\cref{fig:vc-nesting}(b), if the second handler effects a change in \rax, this influences the execution of the first \vc handler that uses the modified \rax. 
In our experiments, we were able to nest to at least a depth of 3.
We report that the hypervisor can also inject consecutive nested  interrupts to change various parts of the \vc handler (e.g., we inject 2 consecutive nested interrupts of depth 1  in ~\cref{fig:vc-nesting}(b)).
There is practically no limit to how many consecutive nested \vc{s} the hypervisor can inject.

\noindent{\bf Nesting \vc in critical section.}
The \vc handler implementation has one critical section. In particular, the hypervisor and the guest VM 
communicate via two shared buffers as part of the GHCB. 
For correctness, in the case that the hypervisor is benign, 
the handler synchronizes accesses to these buffers to avoid race conditions with the \cvm. This creates a critical section in the handler. In typical interrupt handlers, it is standard to disable interrupts when executing a critical section. 
In the case of the \vc handler, to our surprise, we find that the hypervisor can inject nested \vc{s} even in the  critical section execution. In our experiments, we 
were able to achieve a nesting depth of 1 in the critical section for Linux kernel implementation. 
We investigated if this was an implementation bug or an intentional design choice. Our analysis shows that this is a necessary functionality to handle a particular case where both the operands of a \mov operations are memory addresses. Interested readers can refer to \cref{appx:nesting-analysis} for details.

\new{In summary, there are several cases in the VC handler that we can leverage to change or read registers and \cvm memory. The ability to chain and nest \vc{s} allows us to achieve a cascading effect akin to return-oriented-programming (ROP).}

\subsection{Challenges \& decisions}
\label{ssec:challenges}
\new{If the hypervisor can corrupt memory and registers of the \cvm, then we can build powerful attacks. 
Next, we outline the challenges in achieving our goal, especially when the hypervisor has limited access to the \cvm. 
We identify the best ways to use \vc to mount our attacks and provide rationale for our choices of what to exploit. 
}

\noindent{\bf Challenge 1: Targeted \vc injection.} 
To perform a meaningful attack using \vc, we first need to identify instructions in target programs that would result in a meaningful effect as a result of the \vc. 
In our example from~\cref{ssec:intfail-attack}, changing the value of \rax on Line 3 in~\cref{lst:example} leads to the attacker successfully authenticating  the target program. 
Once we have identified the target instruction, we should time the \vc such that it is injected just before our target instruction is executed. 
There are two nuances that we need to address.
First, when we trick the VC handler into accessing certain pages, it might cause stage-1 or stage-2 page faults. 
If a stage-1 page fault occurs during VC handling it will crash the VM.
Therefore, \codename has to ensure that this never happens.
In contrast, stage-2 page faults are not an issue because the hypervisor can always ensure that the pages with the addresses of interest \new{are paged in when the VC handler executes}. 
Second, some pages in the kernel have limited permissions (e.g., \text section is not writable). 
If our attack attempts to write to these pages, it will cause a page fault and crash the VM. 
As we will show in \cref{ssec:code-injection}, \codename gets around this limitation by changing the page permissions before it triggers an access to such a page, avoiding a crash.

\noindent{\bf Challenge 2: Constructing primitives using handler effects.}
The VC handler performs different checks and induces different side-effects depending on the \exitreason.
For example, the \vmmcall handler checks the return value of the \texttt{perform\_VMEXIT} function. 
This function looks up the \exitreason and then performs a series of checks. If the checks pass, it returns OK. 
For \vmmcall, the VC handler checks that the hypervisor has written a value into \rax of the incoming GHCB. 
Therefore, this reason is controlled by the hypervisor making it easy to get around the check in Line $3$ in \cref{fig:vc-nesting} (b).
However, there might be other checks in the VC handler that access protected memory which cannot be influenced by the hypervisor.
For example, the handler looks up the last instruction that was executed and reads the value of the register used as an operand to that instruction. 
The last instruction and the value of its operand are stored in the application context that is protected and cannot be controlled by the hypervisor. 
Such checks could crash the handler, thwarting our attack. 
\new{Therefore, to build \codename primitives we should carefully choose and chain the handler effects to avoid such crashes.}

\noindent{\bf Insight 1: Using only two \exitreason{s}.}
While there are more expressive effects of the VC handler (e.g., \texttt{rdpmc}, \texttt{cpuid}), we find that the handling of intercepts for \vmmcall and MMIO is sufficient to build powerful attack primitives. 
Further, handling these intercepts does not have many side-effects (e.g., changes to memory, changes to registers) and checks that would otherwise corrupt execution. 
So, they can easily be used to build \codename primitives.

\noindent{\bf Insight 2: Limiting to kernel memory.}
The hypervisor can raise \vc while executing in both the user and kernel space. 
Therefore, the hypervisor can use \vc to leak or tamper both user and kernel space registers and memory. 
However, using \vc to attack user-space applications is more challenging than compromising the kernel execution. 
First, since the VC handler is executed in the kernel space, the memory accessed from the handler should be mapped in the kernel. User-space application memory is not mapped as-is in the kernel and therefore the VC handler cannot use the userspace virtual address to access it. The attacker needs to either perform the address translation using the process's page tables or single-step the victim process's lifecycle to track the virtual to guest physical address (GPA) mapping, such that it can supply the GPA.
Second, determining when to inject the \vc exception for user-space applications is not straightforward. 
For example, in \cref{lst:example} the hypervisor should inject the \vc exception after Line $3$. 
Determining when this instruction is executed in user-space requires a single-stepping primitive at instruction granularity. 
In contrast, targeting the kernel space does not incur such challenges.
In the kernel space, the pages of the \text segment are mapped contiguously during boot and therefore we can easily compute the address and page of the instruction we want to trigger the \vc on.
Therefore, to determine when to inject the \vc exception while executing in the kernel only requires using page faults to profile the pages that are being executed. 
Since the hypervisor controls the stage-2 page tables of the \cvm{s}, profiling with page faults is easy. 
Thus, we focus on building our attack primitives and case studies using kernel space code.

\noindent{\bf Insight 3: Target instructions executed after page fault.}
\begin{figure}
    \centering
    \includegraphics[scale=0.7]{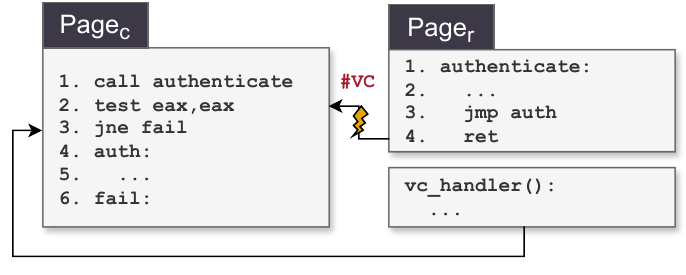}
    \caption{Timing the \vc injection. 1. Hypervisor observes page fault sequence [\pagec, \pager, \pagec] and injects \vc when Line 4 returns. 2. This triggers the execution of the VC handler, which when done returns back to the application on \pagec}
    \label{fig:timing-vc}
\end{figure}
In SEV, the hypervisor manages the stage-2 page tables for \cvm{s} and handles page-faults.
We can use this mechanism to induce page faults by marking pages as non-executable in the \cvm to observe the pages that are executed.
We use this page tracing technique to time the \vc injection. 
First, we limit our attack primitives to target instructions that are executed when jumping or returning from another page (e.g., a call instruction that returns execution from another page, the target of a \texttt{jmp} instruction).
Therefore, we can use the page tracing technique to detect when the execution returns to our target page and inject a \vc.
For example, consider a target program that performs authentication by calling an \texttt{authenticate} function and then tests its return value (see~\cref{fig:timing-vc}).
The hypervisor can use \vc with \vmmcall to change the value of \rax as explained in \cref{ssec:intfail-attack}. 
Therefore, to corrupt the return value of the \texttt{authenticate} function we should target our \vc injection to when the function returns i.e., on the line after the \texttt{call} instruction. 
To identify when the \texttt{authenticate} function returns, the hypervisor marks pages \pagec containing the target instruction and \pager containing a return to the target instruction as not executable. 
Then, when the hypervisor observes a page-fault trace of the pattern [\pagec, \pager, \pagec{}], it knows that this is the return from the \texttt{authenticate} function. 
Therefore, before resuming execution of the target program by marking \pagec as executable again, the hypervisor injects the \vc. 
While this choice limits the instructions that \codename can target, we show that it is sufficient to build powerful primitives and mount attacks on \cvm{s}.

\noindent{\bf ASLR.} 
The kernel uses different Address Space Layout Randomization (ASLR) to thwart attacks that rely on deterministic addresses. 
To use our page-tracing technique and build attack primitives that target specific addresses we need to break these address randomization techniques. 
We use insights from previous works that have explained techniques to defeat the kernel's physical ASLR. This does not require \vc or other \codename primitives.
We use \codename primitives to defeat virtual address space ASLR (see \cref{ssec:virtual-aslr}).

\subsection{Threat model}
We assume an untrusted hypervisor creates and launches AMD SEV-SNP VMs.
The trusted hardware generates an attestation report.
AMD SEV-SNP's trusted hardware ensures that all memory and registers of the \cvm are protected and encrypted. 
Further, on context switches the trusted hardware saves and restores the context of \cvm{s}. 
It also ensures that all register states are cleared before resuming execution of other untrusted code. 
We assume that all trusted software in the \cvm, including the VC handler, and hardware are free from bugs. 
To communicate with the hypervisor, the \cvm sets up shared-memory (GHCB) according to AMD specification. 
We assume that the \cvm uses the GHCB strictly in accordance with the AMD specification and is bug free. 
The hypervisor is still responsible for interrupt delivery and memory management for the \cvm{s} according to AMD specifications, and can observe page faults. 
We assume that all cryptographic algorithms used by AMD SEV-SNP are secure. 
We do not assume any other architectural, microarchitectural, power, or voltage side-channels. 
\section{\codename Primitives}
We build basic primitives either by using one \vc's effects or by cascading \vc{s}. Our main challenge is to use the \vc to induce the desired effect while avoiding the \vc's undesired checks and effects. For example, using \exitreason as \vmmcall and writing to \rax requires passing certain checks as explained in~\cref{ssec:challenges}. 
First, we build a primitive to skip instructions using the \vc handler's effect (\cref{ssec:skip-inst}). 
Next, we  observe that the \vc handler leaks \rax and writes to \rax when \exitreason is set to \vmmcall. 
With these effects we build primitives that read and write to \rax (\cref{ssec:rw-rax}). 
Further, we use these primitives to induce desired effects (e.g., change \rax to an attacker controlled value) and avoid undesirable effects (e.g., skip checks that would otherwise jump to a \texttt{fail} block) while building more powerful primitives. 
Further, we build primitives to read and write to kernel memory using MMIO read and write handling effects in the \vc handler (\cref{ssec:read-mem} and \cref{ssec:write-mem}).
Finally, we show how \codename uses these primitives to inject arbitrary code into the kernel (\cref{ssec:code-injection}). 

\subsection{Skipping Instructions}
\label{ssec:skip-inst}
We build a basic primitive \skipg that uses \vc to skip over an arbitrary instruction during the \cvm's execution. Such a primitive can be used to bypass checks and negate undesired effects in a target program (e.g., to skip over an instruction that jumps to a \texttt{fail} block). 

\noindent{\bf Skip one instruction.}
During normal operation, the CPU does not move the instruction pointer past the instruction that caused \vc, to give the application an opportunity to retry the instruction if it fails. 
Therefore, in~\cref{fig:vc-exception}(a) the return from the VC handler would result in the \vmmcall instruction executing again on Step 5. 
Therefore, the VC handler always has an effect which advances the instruction pointer.  
Depending on the type of the \vc, the VC handler could have other effects (e.g., updates to registers, writes to memory). 
An attacker can induce the effect in the VC handler that increments \rip and use \vc to build a skip primitive (\skipg) that skips arbitrary instructions. 

\codename should ensure that the skip primitive does not have any other undesired effects (e.g., changes to register values) that can corrupt the target application's execution. 
Therefore, we use a \vc with \exitreason set to \vmmcall to build this primitive. 
Handling the \vmmcall has only one other effect besides incrementing \rip; it leaks the values of \rax to the hypervisor and writes a value from the hypervisor to \rax as shown in~\cref{fig:rw-rax} (Lines $3$, $6$). 
Because the \vmmcall handling writes the value from the hypervisor to \rax, the attacker can always ensure that this effect does not corrupt any state in the application. 
Specifically, when the VC handler exits to the hypervisor, the hypervisor first reads the value of \rax from the GHCB. 
Then, it copies this value back to the GHCB to be read by the VC handler for \vmmcall. 
This ensures that the value in \rax remains unchanged. 
Next, the VC handler only increments the \rip if the call from the hypervisor was successful i.e., for \vmmcall the hypervisor wrote a value into the \rax field of the GHCB (Line $11$ in~\cref{fig:rw-rax}). 
Therefore, for \vmmcall this is fully controlled by the hypervisor.  
Specifically, by copying the value of \rax into the GHCB, the hypervisor also ensures that the function on Line $4$ always returns OK.
The VC handler does not increment the \rip by a fixed number of bytes. 
Instead, it looks up the last instruction that was executed from the program context. 
Then, it computes the size of that instruction and the number of bytes to increment (Line $10$) ensuring that a full instruction is always skipped.

\noindent{\bf VC chaining: Skip $n$ instructions.}
The hypervisor can chain the skip primitive to skip any number of consecutive instructions. 
For this, it pauses the execution of the target program by marking the page that the VC handler returns to as non-executable. 
This ensures that every time the VC handler returns to the target program, a page fault is generated. 
The hypervisor then uses the skip primitive to skip 1 instruction on each page fault of the program. 
The VC handler looks up the current instruction that \rip points to (Line $10$ in~\cref{fig:rw-rax}), which ensures that each subsequent instruction is skipped correctly irrespective of its length.

\subsection{\bf Read \& Write \rax.}
\label{ssec:rw-rax}
Using malicious \vc{s}, we build \codename primitives to read and write to \rax. Return values from function calls are stored in \rax which makes it particularly interesting. 
Therefore, leaking or tampering the value of \rax can be used to construct powerful attacks. 

\begin{figure}
    \centering
    \includegraphics[scale=0.67]{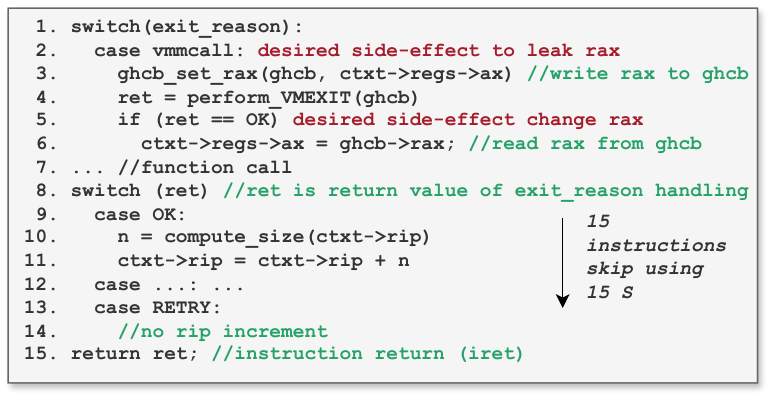}
    \caption{Psuedo-code of VC handler with effects.}
    \label{fig:rw-rax}
\end{figure}

\noindent{\bf Read \& Write \rax and skip $1$ instruction.}
As explained in~\cref{ssec:intfail-attack}, a malicious hypervisor can use the \vc to 
read and write the \rax register at any point during the instruction execution. 
Further, as discussed in~\cref{ssec:skip-inst}, the VC handler has an effect of incrementing the \rip. 
We build attack primitives \raxrskip and \raxwskip that read and write to \rax respectively and then skip $1$ instruction.  
To build these two primitives, we simply inject one \vc with \exitreason as \vmmcall. 
This will leak the value of the target program's \rax (Line $3$ in~\cref{fig:rw-rax}) and copy a value from the hypervisor into the \rax of the target program (Line $6$~\cref{fig:rw-rax}) and skip one instruction.
Even though we define \raxwskip as a separate primitive, the VC handler always leaks the value of \rax (Line $3$ in~\cref{fig:rw-rax}) and therefore \raxwskip implicitly always reads \rax as well.

\noindent{\bf Read \& Write \rax without skipping.} 
The effect of incrementing the \rip in \raxrskip and \raxwskip is sometimes not desirable while building attacks where we want to execute the instruction that we injected the \vc on. 
To negate the \rip increment effect we build $2$ more primitives \raxr and \raxw that read and write to \rax respectively without skipping an instruction. 
Negating the undesired \rip increment effect is not straightforward.
Notice that the \rip is not incremented if the return value on Line $4$ in~\cref{fig:rw-rax} is RETRY (Line $13$).
However, the hypervisor cannot force this condition on Line $4$ for \vmmcall handling. 
Instead, to negate the instruction skipping effect, we construct a mechanism that skips all instructions ($15$ instructions) starting from Line $8$ to Line $13$. 
This ensures that the VC handler executes the RETRY case even though the return value on Line $4$ was OK. 
\new{With this, the desired effect on Line $3$ for \raxr and on Line $6$ for \raxw is preserved and the undesired effect on Line $11$ is negated.} 
To skip these instructions, we use a single-level of nesting using the skip primitive (\skipg).
We chain the skip primitive $15$ times in succession. Therefore, (\raxrskip).15\skipg $\equiv$ \raxr and (\raxwskip).15\skipg $\equiv$ \raxw. 
This ensures that the \rip is not incremented and the VC handler simply returns back to the target program. 
\new{To decide when to inject the first \skipg primitive in the chain, we notice 
that the VC handler performs a call to a function (Line $7$) before our target instruction on Line $8$. 
Therefore, we use our page-trace mechanism to inject the first \skipg on the return of this function call. }

\begin{figure*}[h]
    \centering
    \includegraphics[scale=0.65]{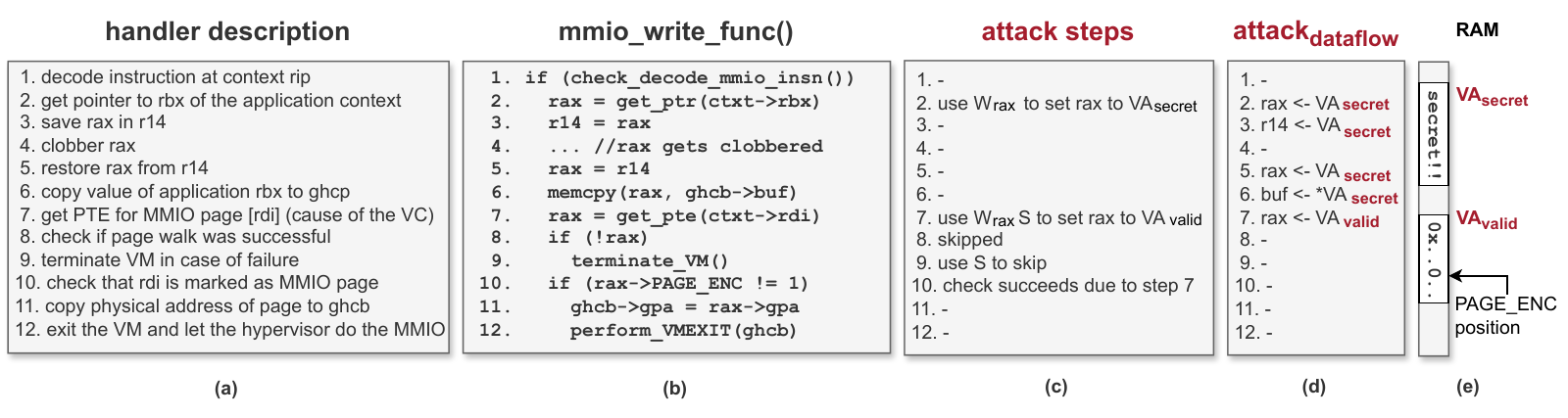}
    \caption{Read memory primitive. (a) Handler description for \vc caused by instruction \texttt{mov [rdi], rbx}. (b) Pseudo-code of MMIO write handling. (c) Attack steps. (d) Attack dataflow. (e) Memory used for attack. }
    \label{fig:read-gadget}
\end{figure*}

\subsection{Reading kernel memory}
\label{ssec:read-mem}
The attacker can use \vc{s} to read values from any kernel memory (\readg). 
To successfully read any kernel memory of the attacker's
choosing, the attacker needs $2$ capabilities: (a) control a pointer to kernel memory, and (b)  dereference the attacker-controlled pointer and copy the value to the GHCB. 
We observe that the attacker can use the VC handler's MMIO write case to gain these capabilities. 
During normal operation, the CPU invokes the \vc for MMIO write when a \mov instruction copies a register value to a memory-mapped region (e.g., \verb|mov [rdi], rbx| where \texttt{[rdi]} points to a memory-mapped region). 
To enable the hypervisor to perform the actual MMIO write, the handler copies the value of the target program's register into the GHCB as shown in ~\cref{fig:read-gadget} (Line 6).   
To perform the memory copy, the handler first fetches a pointer to the program's context on stack (Line 2) and then de-references it (i.e., \texttt{memcpy} dereferences \rax). 
Our main intuition is that we can use the primitives that write to \rax to gain control of the pointer in the VC handler before it is dereferenced. When the pointer is dereferenced, the VC handling for MMIO writes will copy a value from the attacker-controlled pointer to the GHCB. 
This satisfies the $2$ capabilities the attacker needs to successfully read kernel memory. 
Next, we explain the benign MMIO write case in the VC handler shown in in~\cref{fig:read-gadget} (a) and (b).

\noindent{\bf Benign execution of VC handler for MMIO write.}
To ensure that the hypervisor can perform the MMIO write operation correctly, the VC handler's MMIO write case copies the value of the register to write and the guest physical address of the MMIO region to the GHCB. 
To determine the register and the region of memory, the handler first decodes the last instruction that was executed in the program's context which caused the \vc (Line $1$ in~\cref{fig:read-gadget}).
The handler has these checks because it expects the instruction that caused the \vc to be a \mov instruction that performs a write to memory. 
The code-snippet in~\cref{lst:check-decode-insn} shows this check (Line $3$). 
Further, once the instruction is correctly decoded, it also sets the operands of the instruction (Line $6$) in the VC handler's stack. 
\begin{lstlisting}[language=C, caption={Pseudo-code for check and decode mmio instructions \\ during MMIO write handling}, label={lst:check-decode-insn}]
check_decode_mmio_insn():
  type = decode(ctxt -> insn) 
  if (type != MMIO_WRITE)
    goto fail    
  // sets the instruction operands
  get_insn_operands()
\end{lstlisting}
\new{After determining the operands, the VC handler needs to copy the value of the register to the GHCB. 
When an exception occurs the hardware pushes the context of the program that caused the exception on the exception handling stack.} 
Therefore, to get the value of the register, the VC handler fetches a pointer to the register in the program's context and uses \texttt{memcpy} to copy the value in the register to the GHCB's buffer (Line $6$ in~\cref{fig:read-gadget}(b)). 
Then, to determine the GPA of the memory-mapped region, the handler performs a page-walk to get the page table entry (Line $7$ in~\cref{fig:read-gadget}(b)). 
It checks that the page table entry is valid (Line $8$) and that the memory-mapped region is in plaintext and in hypervisor-accessible memory (Line $10$). 
It finally exits the VM to go to the hypervisor. 
In summary, handling the MMIO write has $2$ side-effects : (a) dereference a pointer to memory and copy its value to the GHCB, and (b) copy a valid GPA of an unencrypted memory region to the GHCB. 

The first effect that dereferences the pointer in memory and copies the value to the GHCB is desirable and can be used to gain the $2$ capabilities that the attacker needs. 
However, for \readg the checks and effects that result from copying a valid GPA to the GHCB need to be negated. 

\noindent{\bf Building the read-memory primitive (\readg). }
To build the read-memory primitive, we start with a \vc with \exitreason set to \mmiowrite which triggers the VC handler for MMIO write. 
We need to ensure that the $2$ capabilities that the attacker needs to read kernel memory are satisfied. 
The attacker should control a pointer while handling MMIO, and this pointer should be dereferenced and copied to the GHCB.
We can achieve this by writing an attacker-controlled pointer address into \rax when the function on Line $2$ in~\cref{fig:read-gadget} returns. 
The attacker-controlled pointer in \rax is then dereferenced by the \texttt{memcpy} on Line $6$ which copies the value ($8$ bytes) to the GHCB buffer. 
The \rax is first saved to \texttt{r14} (Line 3) and then restored (Line 5) before performing the \texttt{memcpy}. 
Therefore, it is important for our primitive that the write to \rax on Line $2$ does not skip the next instruction on Line $3$.
Therefore, we use the \raxw primitive to change the value of \rax when the \texttt{get\_ptr} function returns (Line $2$).
With this we have achieved the desired effect of controlling a pointer to memory, and correctly copying it to the GHCB. 

However, handling the MMIO write has another undesired effect.
It performs a page-walk (Line 7) and checks the page table entry (Line 8, 10) before performing a \vmexit. 
If the page-walk or the checks fail, the handler returns an error and does not perform the \vmexit to the hypervisor. 
Therefore, in our read-memory primitive, we need to ensure that the side-effects and checks are negated and the VC handler sucessfully exits to the hypervisor. 
We observe that the result of the page walk on Line 7 is saved into \rax. 
We can use our primitive that writes to \rax again to change the value of \rax such that it passes the checks on Lines 8 and 10. 
Here, we opt to nest the primitive that writes to \rax and skips $1$ instruction (\raxwskip). 
This implies that the instruction that performs the check on Line 8 is skipped. 
However, to fully bypass the check we should ensure that one more instruction on Line 9 is skipped using the skip primitive (\skipg).
With this, we have successfully bypassed the first check. 
The second check (Line 10) ensures that the memory-mapped region is in unencrypted memory by checking the \texttt{PAGE\_ENC} bit of the page table entry. 
This is a simple memory dereference at an offset from the base address in \rax. 
Therefore, we identify a region of memory in the kernel such that the value at this offset is always $0$. We set it as the address in \rax before \texttt{get\_pte} returns on Line $7$. 
This will guarantee that the check on Line $10$ will pass. 
The \rax is dereferenced once more on Line $11$ to write the GPA to the GHCB buffer. However, for our \readg this value is not important. 
Therefore, MMIO write handling will gracefully exit to the hypervisor. 

\new{~\cref{fig:read-gadget} only shows the pseudo-code of the VC handler for MMIO write. For simplicity, we show the checks on Line $10$ to be inlined after the check in Lines $8$ and $9$. In the Linux implementation, the  check represented on Line $10$ is in a different function. 
Therefore, we cannot use a chain of skip primitives to skip all instructions from Lines $9$--$12$ to perform the 
\vmexit. }
Instead, we detect the return from the functions using our page fault tracing method and then inject subsequent \vc{s} as described above. %

\noindent{\bf Target instruction for \readg.}
We identify a target \mov instruction that writes register values to memory in the scheduling subsystem of the kernel.
This instruction is on the hot path of the kernel's execution and is frequently executed.
Furthermore, this instruction is executed after a \texttt{jmp} from another page.
Therefore, we use the page fault profiling as explained in~\cref{ssec:challenges} to identify the page fault sequence and time the \vc with \exitreason as \mmiowrite to trigger the read-memory primitive (\readg). 

\noindent{\bf Reading n bytes of kernel memory.}
\readg only reads $8$ bytes of memory from the kernel. 
To read more memory, we pause the execution by marking the page of our target instruction as non-executable.
Then on each page fault of our target instruction, we use \readg.
Using this, we can chain \readg to read kernel memory in $8$ byte chunks.

\subsection{Writing to kernel memory}
\label{ssec:write-mem}

We can use \vc to build a primitive that writes arbitrary data to arbitrary addresses in the kernel memory (\writeg). 
For this, the attacker needs $2$ capabilities, similar to the read primitive: (a)  control a pointer to kernel memory, and (b) the attacker-supplied value in the GHCB should be copied to the attacker-controlled pointer. 
We observe that the hypervisor can achieve these capabilities using the MMIO read case. 
During normal operation, this handler is triggered on a \mov instruction that reads a value from memory into a register. 
Therefore, similar to the MMIO write case, the read case has $2$ effects: (a) it fetches a pointer to the program context and then copies the value from GHCB into the register, and (b) it copies a valid GPA to the GHCB. 
This allows the benign hypervisor to read data from the memory-mapped region, send it to the VC handler which then copies the data into the register. 

\begin{figure}
    \centering
    \includegraphics[scale=0.65]{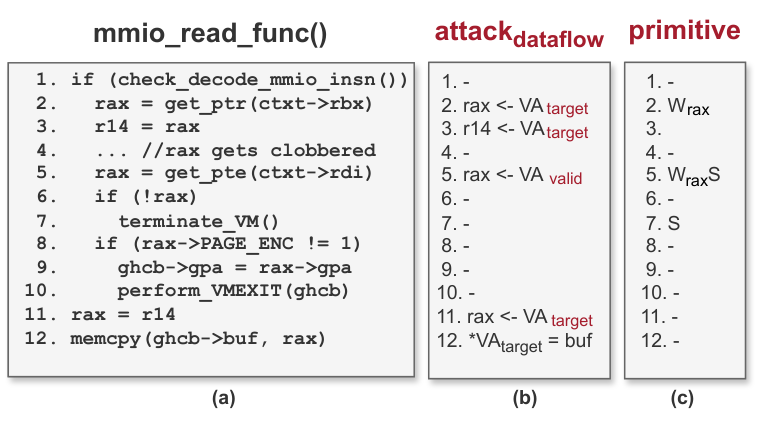}
    \caption{Write memory primitive. (a) Pseudo-code of MMIO read handling. (b) Attack dataflow. (c)\codename primitives used for attack}
    \label{fig:write-gadget}
\end{figure}

To achieve the attacker's objectives of writing to an arbitrary address in the kernel, \codename first uses the \raxw primitive to get hold of an attacker-controlled pointer (Line $2$ in~\cref{fig:write-gadget}). 
This pointer is then dereferenced (on Line $12$) when the handler copies attacker supplied data from the GHCB into the attacker-controlled pointer address. 
Finally, as with \readg, \codename chains \raxwskip and \skipg to negate the undesirable checks and side effects on Lines $6$ and $8$. 
With this, the value from the hypervisor ($8$ bytes) in the GHCB is copied into a kernel memory location that is controlled by the hypervisor.

\noindent{\bf Target instruction for \writeg.}
We identify a target \mov instruction in the kernel's scheduling subsystem, that writes a value from memory to a register and is also the first instruction executed after a return from another page. 
Therefore, we time the injection of \writeg using our page trace mechanism.
Further, to write more than $8$ bytes to kernel memory, we pause the execution of the target instruction by marking the page as non-executable. 
Then we chain \writeg to write to kernel memory in $8$ byte chunks.

\subsection{Arbitrary code injection}
\label{ssec:code-injection}
Using \codename primitives explained so far, we can perform arbitrary code injection in the kernel. 
To inject code to be executed in the kernel, we need to write to the \text section. 
By default, the kernel sets up its page tables such that the \text section is executable but not writable. 
To get around this constraint, we first use our \readg and \writeg primitives to change the permissions in the kernel page tables. 
For this, we first read the value of the kernel's CR3 using the read memory primitive. 
This value indicates the base address address of the kernel's page tables. 
We identify the virtual address of the target page where we want to inject our code in the kernel's \text section.
We then use \readg to perform a page walk starting from the base address to this page. 
At each level of the page table, we use \writeg to change the permission of the page to be writable.
Once all the permissions are changed, our target page is writable. 
Finally, we can write shell code using a chain of \writeg to the target page.  
It remains to show how we execute our injected code in the \cvm. 
This depends on where the code is injected as shown in our case studies (\S~\ref{sec:case-studies}).

\section{Bypassing ASLRs}
\label{ssec:bypassing-aslrs}
\begin{figure}
    \centering
    \includegraphics[scale=0.75]{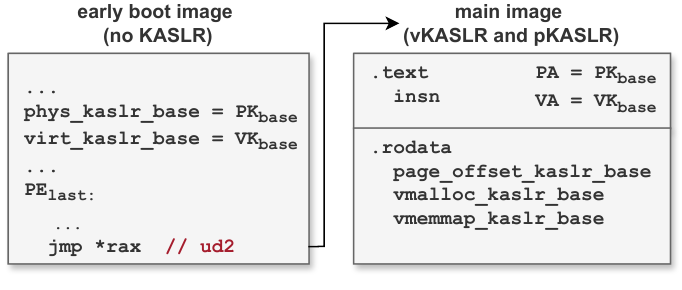}
    \caption{Kernel boot images and KASLR. The red instruction is inserted by the hypervisor for offline profiling.}
    \label{fig:kaslr}
\end{figure}

Address Space Layout Randomization (ASLR) is a simple way to stop attacks that rely on deterministic addresses. 
Linux uses ASLR for both physical and virtual addresses.

\subsection{Physical kernel ASLR (pKASLR)}
The physical ASLR protection in the kernel (pKASLR) randomizes the physical address where the kernel is loaded for every kernel boot. 
Previous works explain techniques to break pKASLR of SEV-ES VMs using page faults observed by the hypervisor~\cite{SEVerity, SEVurity}. 
We use insights from these works to build an attack that compromises the pKASLR of SEV-SNP VMs. 
To compromise pKASLR we do not need any \codename primitives but just the page fault tracing technique in the hypervisor. 

\noindent{\bf Overview.} The pKASLR base is the address of the first page of the main kernel image (\pkbase). 
The kernel memory is allocated contiguously from this base address. 
So, the address of every page in kernel memory is a fixed offset from \pkbase. 
Therefore, to compromise pKASLR we simply need to determine the value of \pkbase. 
To do this, we analyze the Linux kernel's boot process.  
We find that before the main kernel image is loaded, an early boot image sets up \pkbase by choosing a random address~\cite{SEVurity}. 
Then, it loads the main kernel image at this random base address and jumps to it (see~\cref{fig:kaslr}). 
Crucially, this early boot image which sets up the base address is not subject to any ASLR protection. 
Therefore, for every execution of the Linux kernel, the physical addresses of the early boot image remains constant. 
We can use this insight to find \pkbase by profiling the kernel boot.
Specifically, if we determine the address of the last page that the early boot image executes (\pelast), then the very next page that is executed will always be \pkbase (see~\cref{fig:kaslr}). 

\noindent{\bf Attack steps.} To find \pelast, we modify the early boot image and replace the \texttt{jmp *rax} in~\cref{fig:kaslr} with a dummy instruction \texttt{ud2}. 
This step can be performed by the hypervisor offline without the victim's  knowledge. 
We boot this modified kernel image and induce page-faults for each page executed during kernel boot.
This enables us to gather a page trace of all the pages executed during the kernel boot and their corresponding physical addresses. 
Using this page trace we determine the physical address of the last page (\pelast) that is executed by the early boot image.  
Because the early boot image is not subject to ASLR, \pelast is constant across all boots of the Linux kernel. 

Next, we boot the unmodified victim kernel image that contains the jump from the early boot image to the main kernel image (see~\cref{fig:kaslr}). 
Now, the page executed right after \pelast is the physical address of the main kernel image (\pkbase) which we want to de-randomize. 
Therefore, we can deterministically leak the value of \pkbase compromising pKASLR.

\subsection{Virtual Kernel ASLRs with \codename Primitives}
\label{ssec:virtual-aslr}
Like physical ASLR, the Linux kernel randomizes the addresses of its virtual address space using virtual ASLR (vKASLR). 
Once we have compromised pKASLR, we can compute the physical address of any function in the kernel. 
Our main insight to compromise virtual ASLR (vKASLR) in the kernel is to find an instruction during the kernel's execution that loads the virtual address of a kernel's function into \rax. 
At this point, we can use our \raxr primitive, to leak the virtual address of the function. 
We observe that the kernel sets up its virtual address space such that the kernel memory is contiguous in the virtual address space. 
Therefore, we can compute the base of the virtual kernel ASLR (\pvbase) as a fixed offset from the function's virtual address that we leak. 

We analyze the kernel and find an instruction in the \verb|secondary_startup_64| routine that loads the virtual address of the function \verb|x86_64_start_kernel| into \rax (Line $2$ in the code snippet below).
\begin{lstlisting}[language={[x64]Assembler}]
xor  ebp, ebp 
mov  rax, [rip + offset] 
push rax
ret
\end{lstlisting}
We inject \raxw right after \texttt{ret} on Line $4$ to leak the virtual address of \verb|x86_64_start_kernel|. 
Using this, we compute the value of \pvbase as a fixed offset from the \verb|x86_64_start_kernel| function. 

The Linux kernel randomizes regions for the identity map (\texttt{page\_offset}), \texttt{vmalloc}, and \texttt{vmemmap}  separately~\cite{tagbleed}. 
The randomized base address for these regions is stored in the \texttt{.rodata} section that is only subject to vKASLR. 
Therefore, once we have compromised vKASLR, we can compute the virtual addresses of all the base addresses and use \readg to leak their values.

\section{Case studies}
\label{sec:case-studies}
We present $3$ case studies that use \codename primitives. 
Our case studies do not need profiling beyond page faults. 

\subsection{Leaking keys from Kernel TLS}
\label{ssec:ktls}
\begin{figure}
    \centering
    \includegraphics[scale=0.65]{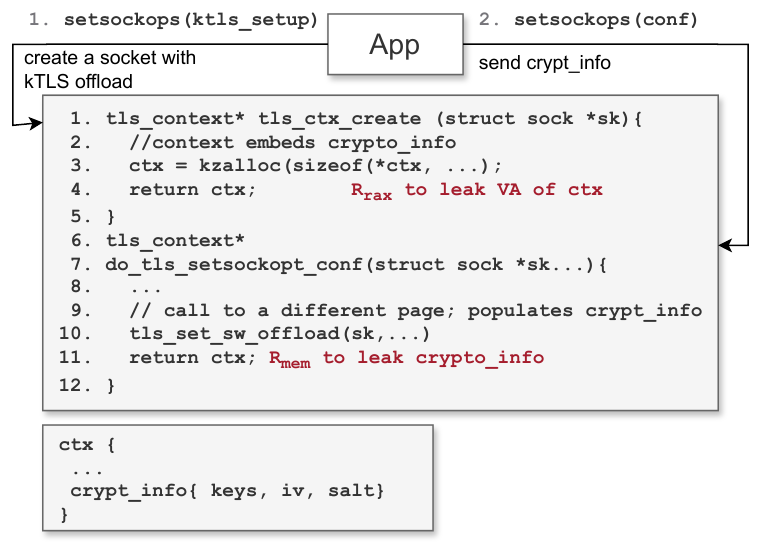}
    \caption{kTLS offload in the Linux kernel.}
    \label{fig:ktls}
\end{figure}
The Linux kernel allows userspace applications to offload TLS computation to the kernel. 
This boosts performance by reducing the number of data copies from user to kernel space. 
To use the kernel TLS offload, the userspace uses the system call \verb|setsockopt| to indicate to the kernel to hook on packets sent through a particular socket (Step 1 in~\cref{fig:ktls}). 
Then, the userspace application performs the TLS handshake to setup symmetric session keys. 
Once the symmetric session keys are setup, the application transfers the keys along with all the information required by the kernel (session key, IV, salt, and sequence number) to encrypt and decrypt the session packets.
The application sends the \verb|struct crypto_info| using the \verb|setsockopt| system call (Step 2) to send this data to the kernel. 
We aim to leak the \verb|crypto_info| using \readg to read this struct such that we can encrypt and decrypt any session packet compromising the TLS session. 

To perform an end-to-end attack, we need to find the virtual address of the \verb|crypto_info| struct for the target socket. 
The userspace first creates the socket and establishes the kernel TLS using \verb|setsockopt| (Step $1$~\cref{fig:ktls}). 
On receiving this system call, the kernel allocates kernel memory for the socket context using \verb|kzalloc| which returns a pointer to the memory in \rax (Line $2$). 
The context structure contains the \verb|crypto_info| struct for the socket as shown in~\cref{fig:ktls}. 
Therefore, we leak the virtual address of the socket's context using the read \rax primitive (\raxr) and compute the virtual address of \verb|crypto_info| as \vaci. 
At this point, the userspace application has not yet transferred the keys and other sensitive information to the kernel.
Therefore, we have to wait for the userspace application to transfer the session keys and other information to the kernel using a second \verb|setsockopt|. 
When the userspace invokes this system call with the values (Step $2$ in~\cref{fig:ktls}) the kernel copies the information with the session keys to \vaci (Line $6$).  
After this, we can leak the \verb|crypto_info| using \readg.
To trigger \readg, we observe that after storing the values into \verb|crypto_info| the system call handling in the kernel calls another function (Line $7$) which is on a different page. 
So, we use the page-fault technique to inject \readg on Line 11 and read out the \verb|crypto_info| struct.
To verify that this information is sufficient to compromise the TLS session, we record all packets for the session and check that we can decrypt them correctly using the leaked information.

\subsection{Disabling Firewall}
\label{ssec:firewall}
The \texttt{iptables} utility allows system administrators to setup packet filter rules for incoming and outgoing traffic to create firewalls using the Linux kernel. 
While the \texttt{iptables} rules are configured in the userspace, the actual packet filtering is performed in the kernel. 
The kernel invokes the \verb|nf_hook_slow| shown in~\cref{lst:nf-hook-slow} for the filtering.
\begin{lstlisting}[language={[x64]Assembler}, caption=Exiting implementation of \texttt{nf\_hook\_slow}, label={lst:nf-hook-slow}]
endbr64
movzwl eax, [rdx]
cmp    ecx, eax        ; ... do packet filtering
\end{lstlisting}
We aim to prevent the kernel from filtering the packets and compromise its firewall protections. 
To perform this attack, we can use \codename's code injection from~\cref{ssec:code-injection} and replace the code of the \verb|nf_hook_slow| function. 
Specifically, we inject shell code showin in ~\cref{lst:firewall-inject} at the virtual address of the \verb|nf_hook_slow| function. 
\begin{lstlisting}[language={[x64]Assembler}, caption=Code to inject in place of \texttt{nf\_hook\_slow}, label={lst:firewall-inject}]
endbr64
mov $1, rax
ret
\end{lstlisting}
Our code simply ensures the function returns success ($1$) without performing any filtering. 
We have to always inject $8$ bytes at a time because we use the \writeg primitive. 
Therefore, while it would have been sufficient to inject Lines $2$ and $3$ to achieve our objective, we inject Line $1$ to satisfy the alignment of $8$ bytes.
We inject all $3$ lines of shell code as 16 bytes using 2 \writeg.
To verify that our code injection works as expected and prevents the kernel from performing the packet filtering, we setup a victim \cvm with \texttt{iptable} rules that drop all incoming and outgoing packets in the VM.
This implies that the firewall blocks all network traffic to the \cvm. 
Then, we launch our attack, inject the shell code, and send ICMP packets to the VM from the hypervisor and observe that we receive responses from the SEV VM. 
This shows that our code injection was successful, and any firewall protection setup using \texttt{iptable} rules are bypassed in the \cvm. 

\subsection{Gaining a Root Shell in the SEV-SNP VM}
\label{ssec:root-shell}
The kernel exposes an API called \usermodehelper to kernel modules.
This API allows kernel modules to start a user space application from the kernel context.
This API takes as arguments an executable to run in the user space along with arguments and environment variables for the executable.  
When invoked by a kernel  module, it starts a process with the executable in user space with root privileges.
For example, executing the command below from \usermodehelper spawns a root shell in user space on the \cvm. This shell takes as input all network data from port $8001$. Therefore, if executed, this command allows all unauthorized remote adversaries root access to the \cvm. 
\begin{lstlisting}[numbers=none, language=bash]
/bin/bash -c rm /tmp/t; mknod /tmp/t p; 
/bin/sh 0</tmp/t \| nc -ln 8001 1>/tmp/t
\end{lstlisting}
To attack a victim \cvm, we aim to maliciously trick it into executing \usermodehelper with this command as the argument. 
To do this, we generate shell code that executes this command using the \usermodehelper API. 
We choose to inject this shell code in the Linux kernel's function that receives and handles ICMP packets (\verb|icmp_rcv|). 
To inject the shell code we compute the virtual address of the \verb|icmp_rcv| function and use the \codename code-injection from~\cref{ssec:code-injection} which overwrites the function's code. 
Now, we send an ICMP packet to trigger our shell code in the SEV VM. 
When the shell code calls the \usermodehelper API, it creates a new userspace process with root privileges that provides a root shell that listens on port $8001$. 
Then, to interact with the spawned shell we connect to the VM from the hypervisor using \texttt{netcat}. 
There is no authentication or firewall, so our attack connection succeeds and we get a root shell using \codename. 

At this point, we can not only leak all private keys on the file system (e.g., SSH private keys) but also install keys of our choice, manipulate userspace programs, and corrupt stored data~\cite{sev-extract-sectrets, sev-step, cachewarp, SEVerESt}. While \codename does not attack user space by choice (\cref{ssec:challenges}), it achieves the same goals by compromising the kernel space.

\section{Implementation \& Evaluation}
\noindent{\bf Experimental Setup.} 
We perform our experiments on an AMD EPYC 9124 16-core 3.7 GHz processor with 192GiB RAM with SEV-SNP Gen 4 with microcode version~\texttt{0x0a10113e}. 
We boot Linux kernel v6.5.0~\cite{snp-host-latest-tree-ad9c0bf475} with Ubuntu 20.04.6 LTS as guest using QEMU v8.0.0.
On the host we use the same kernel with Ubuntu 23.10 as disk-image.
For the rest of this section, we report measurements averaged over $3$ experiments. 

\noindent{\bf Injecting \vc.}
To inject the \vc with \exitreason for \codename we modified the KVM subsytem of the host kernel with $17$ \loc.
We inject multiple \vc with \exitreason set to \vmmcall and measure the time. 
One round-trip from the hypervisor takes $3.19$  $\mu$seconds on average.

\noindent{\bf \codename primitives.}
We first build our page tracing mechanism by exposing an API in the host kernel to interact with its stage-2 page table management system with $338$~\loc, similar to SEV-STEP~\cite{sev-step}.
Our changes allow us to mark pages as not executable and disable huge pages in the stage-2 page tables.  
Then, we implement a function in the host Linux kernel that uses the page-fault trace to detect specific page-fault patterns and exits from the VM and determines when to use \codename primitives. 
\codename needs to inject $1$, $34$, and $34$ \vc{s} for \skipg, \readg, and \writeg primitives respectively. 
Further, \codename can perform on average $216450$ instruction skips/s, $9.25$ memory reads/s, and $8.95$ memory writes/s.

\noindent{\bf ASLR.}
To break physical KASLR we profile the boot stage and record on average $568960$ page faults for the early boot image with the last page address as \texttt{0x7a753000}. 
To compromise vKASLR we use one \vc for one \raxr.

\begin{table}[t]
\centering
\caption{Case studies.}
\label{tab:case-studies}
\resizebox{1\columnwidth}{!}{%
\begin{tabular}{@{}lrrrr@{}}
\toprule
Case study &
   \begin{tabular}[c]{@{}l@{}}no. of \readg \end{tabular} &
  \begin{tabular}[c]{@{}l@{}}no. of \writeg \end{tabular} &
  \begin{tabular}[c]{@{}l@{}}no. of \vc\end{tabular} &
  \begin{tabular}[c]{@{}l@{}}no. of page faults\end{tabular} \\ \midrule
kTLS       & 8 & 0 & 288  &     2115     \\
firewalls   & 4 & 5 & 238  & 1474    \\
root shell & 4 & 52 & 2891 & 7796 \\ \bottomrule
\end{tabular}%
}
\end{table}

\noindent{\bf kTLS.}
To leak session keys from kernel TLS we use the \codename primitives to implement a host kernel module (+220 \loc) and modify the host kernel (+428 \loc).
We compile NGINX v1.25.3 with kTLS support in OpenSSL v3.2.0 and host a website in the SEV VM. We then access the website from a remote machine on the same network.
Client and server use a standard TLSv1.3 connection with a standard and secure cipher suite TLS\_AES\_256\_GCM\_SHA384 for communication.
We capture the packets for this connection using Wireshark. 
We write a utility to manually decrypt the captured packets in nodejs using \verb|crypto_info| that we leak from the SEV VM as explained in~\cref{ssec:ktls}.
From our experiments, we see that \codename needs 5.97 seconds to recover the full \verb|crypto_info| from when the client offloads the connection to kTLS to when we leak the session key. Further, \codename uses $1$ \raxr and $8$ \readg to leak \verb|crypto_info|.

\noindent{\bf Firewall.}
We manually craft $16$ bytes of shell code that we inject to the \cvm.
Recall that the page we target with our code injection is in the \text section and does not have write permissions set in the page table. 
To edit the page permissions, we perform 3-level page walks and use \writeg to change the permissions which takes on average $2.2$ seconds. 
For both our case studies which perform code-injection, our target address is always in a $2$MiB page. 
Therefore, we only do a 3-level page walk to the page middle directory (pmd) even though the kernel uses 4-level paging.  
Then, we inject our shell code to the target page which takes $0.36$ seconds on average. 
We setup the \cvm with firewall rules that block all network traffic. 
Finally, to test our shell code we use the \texttt{ping} utility in the host to send ICMP packets to the \cvm and observe that we receive responses indicating that our attack was successful. 

\noindent{\bf Root shell.}
To get a root shell on the SEV VM, we use GCC and objdump to generate $392$ bytes of shell code.
After identifying the virtual address of the \verb|icmp_rcv| function, we perform  a 3-level page walk that takes  $3.1$ seconds on average. 
Injecting this shell code takes  $8.1$ seconds on average.
This shell code is significantly larger than what we needed to disable the firewall and therefore takes longer to inject. 
To trigger our shell code, the execution should jump to the \verb|icmp_rcv| function. To do this, we use the \texttt{ping} utility to send an ICMP packet to the 
\cvm. 
Then we use a \texttt{netcat} client to connect to and use the root shell. 

\noindent{\bf Accuracy.}
To determine when to inject \vc, we find functions that are on different pages (see~\cref{ssec:challenges}). 
This technique ensures that our injections always work and the \vc handler is executed as expected. 
However, this function alignment can change with different kernel versions.
For the other versions, an attacker would either need to find different target functions or use single-stepping to target the \vc~\cite{heckler-usenix, sev-step}. 
In this case, the accuracy of the attack depends on the newly identified functions or the single-stepping framework.

\section{Impact Beyond Linux}

Besides Linux, we analyze four open source \vc implementations in different projects (Enarx~\cite{enarx-gh}, Oak~\cite{oak-gh}, Coconut SVSM~\cite{coconut-gh}, and Mushroom~\cite{mushroom-gh}) and two closed source OSes with AMD SEV-SNP support from Microsoft. 

\subsection{Open Source Implementations}
All four open source projects implement at most $3$ out of the total $19$ \texttt{exit\_reasons} listed in~\cref{tab:vc-handler}. Thus their attack surface is smaller than the Linux \vc handler.

\noindent{\bf Coconut SVSM} is the official implementation recommended by AMD that is supposed to be executed in VMPL0. 
As of 4th April 2024, it only supports $3$ \texttt{exit\_reasons}:  \texttt{TRAP}, \texttt{CPUID}, and \texttt{IOIO}.
While these implementations have side effects on the execution state, Coconut SVSM  decodes the instruction that causes the \vc and faults if it is not a valid instruction that can potentially raise a \vc (e.g., \texttt{cpuid}).
Thus, the attacker can only inject a \vc when the victim is executing an instruction to ensure that the instruction decoding succeeds.
Since the instruction decoding and the \texttt{error\_code} are not linked, one can launch \codename attacks.
In our experiments, we could inject a \texttt{CPUID} exception in a location where the processor would normally raise an \texttt{IOIO} exception.
This way we were able to corrupt the register state of the application.

\noindent{\bf Enarx} only implements the \texttt{CPUID} exception.
The handler decodes the \vc generating instruction and compares it to the \texttt{cpuid} opcode.
If these do not match, it terminates the VM.
Thus, we conclude that the Enarx is not vulnerable.

\noindent{\bf Oak} only handles \texttt{CPUID} exception, similar to Enarx.
However, Oak does not decode the \vc generating instruction and will always execute the handler and increment \texttt{rip} by 2.
The attacker can use this to selectively skip instructions or jump in the middle of an instruction. 
Since Oak is not officially supported on QEMU and is mainly used with Google's internal hypervisor we were not able to test our exploit. However, we privately reported this to Oak maintainers on 15th March 2024, they acknowledged our attack and patched the handler~\cite{oak-gh-fix}.

\noindent{\bf Mushroom} implements a custom VMPL0 and VMPL3 kernel. 
It supports restricted injection and fetches the \texttt{error\_code} for the \vc directly from the guests VMPL3 VMCB.
Since the hypervisor cannot control the \texttt{error\_code}, Mushroom is not vulnerable to \codename as long as the VMPL3 kernel uses alternate injection mode (explained in \cref{sec:hw-defence}). 

\subsection{Closed Source Implementations}

We boot two Windows VMs using images with SNP support: (a) Windows Server 2022 Datacenter evaluation ISO on VirtualBox in a local setup; and (b) Windows Server 2019 Datacenter VM on Azure machine with SNP support.
We extract the kernel and system files on both these setups. 
The kernel includes symbols, but no reference to SNP-specific terms such as ``sev'', ``ghcb'', ``snp''. Perhaps the SNP functionality is in a loadable module. 
However, C:/Windows contains thousands of files. Our file search for SEV terms returns zero hits.
Instead of static code search, an alternative is to dynamically observe the execution from the hypervisor. 
KVM lacks support to boot Windows in SNP mode. HyperV supports SNP but is closed-source, so we cannot easily change it to inject \vc.
In summary, analyzing Windows internals proves exceptionally challenging and we were unable to check whether Windows implements the \vc handler or supports restricted and alternate injection modes.

\section{Potential Mitigations}

We propose defenses to detect \codename and stop the hypervisor from injection external \vc{s}.

\subsection{Software-based Defenses}
\label{ssec:sw-defence}

The defense can augment the VC handler. First, it checks the instruction that caused the \vc.  
Then, it can determine if the \vc was raised due to a valid instruction that can be intercepted, before processing it. 
For example, consider that the hypervisor triggers a \vc with \exitreason as \vmmcall on a \texttt{cmp} instruction for the application in \cref{lst:example}. The above defense will prevent the \vc handler from executing the \vmmcall handling logic and corrupting \rax. 
Instead, the handler will decode the instruction as \texttt{cmp}, see that it is not a legitimate instruction that should be intercepted, and discard the \vc. 
Further, the handler must use the instruction it decodes rather than the \exitreason register that can be controlled by the hypervisor. 

The decoding approach is sufficient for most intercepted instructions because they have distinct opcodes (e.g., \vmmcall, \texttt{rdtsc}).
However, some instruction intercepts (e.g., MMIO, reads and writes to debug control registers) are triggered on a \mov instruction.
For these cases, instruction decoding and checking against the valid list of instructions is insufficient. 
The VC handler must also check that the \mov instruction has the correct operands (e.g., memory-mapped regions, debug register \texttt{dr7}), thus complicating the logic.
For example, the VC handler has to perform page-walks to see if the address accessed by the \mov instruction was in the memory-mapped region. 
To complicate matters further, instruction intercepts for alignment checks could be triggered by any instruction (e.g., accesses to unaligned memory). 
So the VC handler will have to check, for every instruction, if it accessed unaligned memory. 

When the exception handler performs the above checks, it has to mask all interrupts, which incurs overheads. 
More importantly, ensuring that the checks are complete is challenging and error-prone. 
In fact, as explained in~\cref{ssec:read-mem}, the VC handler already checks and decodes the instruction for MMIO handling. 
However, in light of \codename, the current decoding logic is (a) incorrect: it treats simple register accesses as accesses to memory; (b) incomplete: it does not check that the address accessed by the \mov is actually memory-mapped, just that the access was to an unencrypted region. 
Therefore, this logic is insufficient or incorrect to securely determine if the \vc was caused by an MMIO operation. 
In summary, a purely software-based defense does not guarantee security and can be potentially bypassed. 
\noindent{\bf Linux Patch for \codename.}
AMD implemented this software patch as a quick fix to prevent arbitrary code execution~\cite{linux-vc-patch}.
We worked with the AMD team and improved the initial version of the patch to also cover the early boot image.
Even with this patch, an attacker can still leak $8$ bytes of the \cvm. 
The guest copies the register content to the GHCB and validates the legitimacy of the VC next.
The hypervisor can infer the copied $8$ bytes even if the guest terminates itself due to an illegitimate \vc. 
This serves as a case in point for the incompleteness of software defenses.

\subsection{Hardware-based Defenses}
\label{sec:hw-defence}

There are two main approaches to defend against \codename that both require hardware support to either protect the malicious hypervisor from tampering \exitreason from or injecting \vc.

\noindent{\bf Protecting \exitreason.}
One key requirement for \codename is the ability to 
set the \exitreason. 
An obvious solution is to protect the \exitreason register, such that the hypervisor can no longer write to it. 
Then, the current VC handler can simply trust the values in the \exitreason and handle them without worrying about the hypervisor corrupting the value. 
This would eliminate the need for complex decoding and checks proposed in~\cref{ssec:sw-defence}.
However, this might break functionality and needs microcode changes.

\noindent{\bf Blocking external \vc injection.}
There are three ways to block the hypervisor from externally injecting \vc into a victim VM's CPU.
First, since there is no valid use-case where a hypervisor needs to inject a \vc, the hardware/microcode can directly block all external \vc injections into SEV VMs. 
Second, AMD SEV supports restricted and alternate injection mode.
If the guest OS uses the Secure VM Service Module (SVSM) feature, it can selectively accept/drop external interrupts~\cite{sev-snp}. 
However, current open-source projects do not fully implement support for these modes.
Third, AMD announced Secure Advanced Virtual Interrupt Controller (sAVIC) on 14 Nov 2023, where the SEV guest OS can mask interrupt vector to selectively allow/drop external interrupts. 
Due to lack of documentation and software, it is unclear if this will mitigate \codename. 

\noindent{\bf Immediate Hardware-defense Adoption.}
All SEV SNP-capable processors support restricted and alternate injection modes which can be used to prevent \codename. 
Currently, the software support to enable these modes are not available in the Linux kernel. 
Efforts by AMD to upstream the secure software support have been hindered with known gaps and problems which are not straightforward to fix. 
For example, interrupts can be injected irrespective of the guest's \texttt{RFLAGS.IF} register state~\cite{rest-inj-patch}.
With these problems, introducing new microcode to block external \vc injection as a hot-fix is the most straightforward solution. 
However, for this AMD has to deem \codename as a hardware vulnerability. 
To the best of our knowledge, AMD considers \codename as a software bug and is working on a combination of restricted and alternate injection to fully mitigate \codename.

\section{Related Work}

\noindent{\bf Attacks and vulnerabilities in SEV VMs.}
Google found multiple vulnerabilities in SEV-SNP by analyzing the security co-processor on AMD CPUs~\cite{google-sev-report}.  
Prior works break SEV by targeting the cryptographic algorithms~\cite{236278}, performing page-remapping in the hypervisor~\cite{SEVered}, and using side channels~\cite{PwrLeaks, 10.1145/3319535.3354216, undeSErVed,Cipherleaks,9833768}.
CacheWarp reverts modified cache lines to corrupt memory writes~\cite{cachewarp}.
Crossline uses hypervisor controlled  address space identifiers (ASIDs) to compromise SEV VMs before they crash~\cite{Crossline}.
Code execution has been demonstrated by exploiting weak memory protection and performing Iago attacks on hypervisor interfaces ~\cite{SEVurity,sev-interface-sec}. 
\codename uses insights, such as page tracing and breaking physical ASLR, 
from these prior works.
All the above attacks target SEV or SEV-ES; except for CipherLeaks~\cite{Cipherleaks} and CacheWarp~\cite{cachewarp} which need 
fine-grained information about the VM to break SEV-SNP.
\codename also breaks AMD SEV-SNP but only needs page fault information.

\noindent{\bf Untrusted interfaces.}
Attacks on Intel SGX exploit system call and other interfaces~\cite{checkoway2013iago,coin,taleoftwoworlds}.
Several defenses build secure interfaces to counter these attacks~\cite{sgx-lock,10.1145/3470534,9155211,6868649}.
For malicious timer interrupts, AEX-Notify proposes a framework for Intel SGX that thwarts attacks that abuse asynchronous exits from the enclaves (e.g., timer interrupts for single-stepping)~\cite{aex-notify}.
TrustZone introduces the concept of secure interrupts to prevent untrusted entities from sending malicious interrupts to protected applications~\cite{TZOS}. 
In their SEV-ES analysis, Radev~\etal and Hetzelt~\etal point that the hypervisor can perform Iago attacks when \vc is used by the VMs to communicate with the hypervisor (e.g., bad rdtsc)~\cite{sev-interface-sec,sev-sec}.
Since they do not consider malicious \vc{s} they conclude that the handler does not leak any sensitive information. 
\codename observes that the hypervisor can trigger \vc at any point and uses  
 benign \vc handlers to compromise the VMs.

\noindent{\bf Attacker's Capability to Profile Victim VM.}
SGX-step uses several known attacks, including APIC timer-interrupts, to build a single-stepping primitive for Intel SGX~\cite{sgx-step}. 
On SEV, the hypervisor can observe page-faults of the guest OS~\cite{SEVered}. 
Cachewarp builds a single-stepping framework for SEV-ES using timer interrupts, information from encrypted register states, and cache timing analysis~\cite{cachewarp}. 
SEV-STEP demonstrates a single-stepping framework for SEV-SNP VMs using page-faults, timer interrupts and eviction set-based cache attacks~\cite{sev-step}. 
\codename only uses page-faults observed by the hypervisor. 
However, if single-stepping techniques are available \codename can mount stronger attacks (e.g., on user space instead of kernel space)~\cite{heckler-usenix}. 

\noindent{\bf Gap when supporting unmodified legacy applications.}
Several prior works run unmodified applications on Intel SGX and Arm TrustZone using abstraction layers (e.g., library OSes) ~\cite{trustshadow, haven, graphene-sgx, occlum, scone, ryoan}.
VM abstractions (e.g., AMD SEV, Intel TDX, Arm CCA) reduce invasive porting changes to legacy applications. 
Running unmodified legacy VMs requires hardware and software support in the kernel (e.g., hypervisor controlled mmio)~\cite{hecate}.
Prior works do not analyze the effects of introducing new interfaces (e.g., \vc) coupled with existing hypervisor capabilities (e.g., ability to inject external interrupts). \codename is a concrete attack in this direction that requires close scrutiny.

\noindent{\bf Kernel Defenses \& Hardening.}
Existing kernel defenses against memory leakage and corruption~\cite{kasan-linux, self-prot-linux}, and code-reuse~\cite{cfi-linux} (e.g., ebpf, system call filtering, seccomp, control-flow-integrity) are orthogonal to \codename. 
We demonstrate \codename on the recommended AMD SEV-SNP kernel, up-to-date patches with all hardening enabled, standard compilers and default configurations recommended by AMD.
The only defenses \codename has to subvert were kernel ASLR and write protection of executable pages.
Stronger defenses, if enabled, may make exploiting \codename harder but may not stop it completely. 
Instead, it calls for systematic and rigorous kernel hardening for TEEs~\cite{tdx-guest-hardening,hardening-linux}.

\noindent{\bf Impact Beyond AMD SEV-SNP.}
Interrupt number 20 corresponds to the Virtualization Exception (\ve) on TDX. The handler is similar to AMD's \vc implementation. 
In theory, \ve is also vulnerable to \codename attack---Linux handler does not decode to check if the VM indeed raised a \ve. 
However, we were unable to exploit the interface for two reasons. Intel TDX prohibits the injection of interrupt number 20 into the guest. Thus we cannot trigger the handler. Secondly, the hypercall used to obtain the register state is served by the trusted TD-module and the untrusted host has no direct way of controlling the arguments. On ARM CCA we were not able to identify a handler with a similar functionality. Thus we conclude that \codename does not apply to Intel TDX and ARM CCA.

\section{Conclusion}

\codename leverages the untrusted hypervisor's ability to inject malicious \vc interrupts into AMD SEV-SNP VMs. \codename triggers the exception handler in the victim VM with well-crafted and well-timed \vc{s} to induce register and memory read/writes as well as arbitrary code injection into the VM memory.
Our three case-studies show that \codename compromises confidentiality and integrity of a victim VM.

\section*{Acknowledgement}

We thank our shepherd, the anonymous reviewers, Mark Kuhne, and Mélisande Zonta-Roudes for their constructive feedback for improving the paper. Thanks to AMD and Borislav Petkov for the mitigation discussions and for developing the Linux patches.

\bibliographystyle{IEEEtran}
\bibliography{IEEEabrv,references}

\appendices

\section{Analysis of Nesting in Critical Section}
\label{appx:nesting-analysis}
x86 ISA supports instructions that operate on multiple memory operands. 
The Linux kernel frequently uses one instruction class that dereferences two memory addresses. 
For instance, the \texttt{movsb} instruction is used for memory copy operations in~\cref{lst:memcpy_linux}.
\begin{lstlisting}[caption=x86 \_\_memcpy in the Linux kernel, label={lst:memcpy_linux}, language={[x64]Assembler}]
SYM_TYPED_FUNC_START(__memcpy)
  mov rax, rdi
  mov rcx, rdx
  rep movsb
  ret
SYM_FUNC_END(__memcpy)
\end{lstlisting}
\texttt{movsb} copies the content from the memory referenced by \texttt{rsi} to the memory referenced by \texttt{rdi}. 
If either of the registers point to a memory-mapped region, MMIO access induced by this instruction triggers the VC handler. 
However, when the CPU passes the \exitreason to the VC exception handler, it does not indicate if the source, destination, or both operands caused the exception. 
Thus, it is up to the handler to detect the faulting memory access. 
As a solution to this problem, the VC handler first reads from the source and then writes the result to the destination.
It effectively splits one optimized \texttt{movsb} instruction into two \texttt{mov} instructions.
When the handler executes the \mov instructions, it expects to raise a nested \vc, due to either one or both instructions.
But it is sure that the instruction causing the nested exception dereferences only one memory address. %
Since the read and write exceptions happen sequentially, Linux must only support a nesting depth of one \vc.
However, the VC handler itself has a critical section (see\cref{lst:enter_exit_crt}).
In the critical section, the VC handler has exclusive access to the GHCB page used for communication with the hypervisor.

\begin{lstlisting}[language=C, caption=VC handler entering and leaving the critical section, label={lst:enter_exit_crt}]
ghcb = __sev_get_ghcb(&state);
...
result = vc_handle_exitcode(ghcb, error_code);
__sev_put_ghcb(&state);
\end{lstlisting}

To support one level \vc nesting in the critical section Linux introduces a second backup GHCB page. 
This is needed in our example because the MMIO \vc originating from a \texttt{movsb} instruction holds a lock to the first GHCB page.
Attempting to handle the read and write accesses sequentially within the first \vc causes the nested \vc{s} to acquire and release the lock of another GHCB page.
As of now the guest kernel only supports two GHCB pages.
This effectively limits the nesting capabilities of \vc{s} in the critical section to one.
If a second nested exception occurs while the previous exceptions are in the critical section, the guest will panic. %
While it is theoretically possible to support the case of a second nested exception in the critical section, there is no benign use-case as of now.
However, nesting limits outside of the critical section are only bounded by the stack size available for the VC handler.

\end{document}